# Left shifting analysis of Human-Autonomous Team interactions to analyse risks of autonomy in high-stakes AI systems

Ben Larwood, Synoptix, blarwood@synoptix.co.uk
Oliver J. Sutton, Synoptix, osutton@synoptix.co.uk
Callum Cockburn, Synoptix, ccockburn@synoptix.co.uk

## Abstract

Developing high-stakes autonomous systems that include Artificial Intelligence (AI) components is complex; the consequences of errors can be catastrophic, yet it is challenging to plan for all operational cases. In stressful scenarios for the human operator, such as short decision-making timescales, the risk of failures is exacerbated. The INCOSE 2035 Vision [INCOSE, 2022] states that Human-centred design with a focus on user experience will be a key factor for success of smart systems. A lack of understanding of AI failure modes obstructs this and so blocks the robust implementation of applications of AI in these systems. This prevents early risk identification, leading to increased time, risk and cost of projects.

A key tenet of Systems Engineering and acquisition engineering is centred around a "left-shift" in test and evaluation activities to earlier in the system lifecycle, to allow for "accelerated delivery of [systems] that work" [C. Arndt et al., 2023]. We argue it is therefore essential that this shift includes the analysis of AI failure cases as part of the design stages of the system life cycle. Our proposed framework enables the early characterisation of risks emerging from human-autonomy teaming (HAT) in operational contexts. The cornerstone of this is a new analysis of AI failure modes, built on the seminal modelling of human-autonomy teams laid out by [LaMonica et al., 2022]. Using the analysis of the interactions between human and autonomous systems and exploring the failure modes within each aspect, our approach provides a way to systematically identify human-AI interactions risks across the operational domain of the system of interest. The understanding of the emergent behaviour enables increased robustness of the system, for which the analysis should be undertaken over the whole scope of its operational design domain. This approach is illustrated through an example use case for an AI assistant supporting a Command & Control (C2) System.

## Introduction

As the capabilities of AI within autonomous systems grow, there is inevitably an increasing desire to use them in high-risk or high-stakes systems (that is, systems that pose severe consequences for errors, a high degree of uncertainty or time-sensitivity, and little chance for effective oversight to reverse errors if they do occur) [Adams, 2001; Deveraux, 2022, Ziegler et al., 2022]. To believe that an autonomous system will "have their back" throughout a high-stakes mission, it has been highlighted that a key trust barrier for operators is the need for transparency, understanding, and accessibility of access to system intent, function, status, and capabilities/limitations [Batarseh et al., 2021; Lyons et al., 2018]. Unlike rigidly predictable systems which can be easily verified by their operators, the presence of features such as AI can unlock new capabilities at the cost of making the system difficult to reason about. Transparency becomes increasingly difficult to analyse for complex human-autonomy teams (HATs), since these teams are promoting cooperation and collaboration over pure one-way direction and instruction (unlike conventional automation) [Lyons et al., 2021]. This change in the nature of the interaction also demonstrates the increasing difficulty of analysing autonomous systems that use AI components – where the increase in autonomous system scope and variability corresponds to a significant increase in the potential for additional uncertainty of system behaviour.

As autonomy increases this also leads to the creation of independent decision loops within HATs. The interactions between these human and machine decision loops form a critical aspect of failures within the team as a whole – where the human-machine interface breaks down. This is also the key area where the operator's intent and the machine's goals can conflict, resulting in emergent behaviour that leads to unpredictable system responses [Miller and Rusnock, 2024]. This is evident in pressured situations – for example, where operators are forced to multitask or task-switch, or when their workload stretches their capability to process and draw conclusions from evidence [Chen et al., 2010].

Previous work in this area [LaMonica et al., 2022] has led to the study of interacting paired loops of Observe, Orient, Decide, Act (OODA) actions, referred as OODA$^2$ diagrams. The human and autonomous system each follow their own OODA loop, possibly at significantly differing rates, and interact to share information. Through analysis of LaMonica et al., two key limitations were identified:

1. Whilst their paper provides a strong foundational method, to deliver enhanced engineering value it needs to form part of a structured analysis, including integration into a defined process and further extensions to allow for the identification of impacts resulting from failures.

2. Although demonstrating instances of specific failures allows for proof-of-concept examples of the methodology, a structured taxonomy that allows for the categorisation of failure modes is needed to support effective AI Assurance mechanisms for complex systems.

In this paper, we set out a generalised and extensible methodology for analysing interactions between a human and an autonomous system in a collaborative team. This is not intended to be exhaustive and should be developed and specialised for individual contexts. It is also focused on failure modes specific to the presence of an autonomous system – it is not trying to replace conventional and established analysis methods such as Human Factors Engineering (for the human/human-machine interface) and Systems Safety Engineering (for the system performance). However, we argue that behavioural modelling around the interactions within a HAT will allow for left-shifting of risk identification and will also support overall system development and key transversal analyses.

## Approach

Our approach may be broadly summarised in five steps: (1) an activity diagram with swimlanes is used to describe the threads of activity and decision making cycles between interoperating subsystems and used to (2) identify specific points of interaction between the human and autonomous system; (3) a generic high-level taxonomy of failure modes is applied to categorise these interactions, which may be (4) further specialised to the specific application being studied; (5) the impact of each failure mode is traced through the activity diagram to identify its upstream mechanisms and downstream effects on the system as a whole. This paper broadly focusses on steps 2 to 4 of this process.

*Identifying interactions in the activity diagram*

Within an activity diagram, points of interaction between distinct (human/autonomous) systems are characterised as directed edges connecting a node relating to an action performed by one system to a node representing an action performed by another system. By structuring the activity diagram as an OODA[2] activity structure, we map the downstream node of an 'interaction' edge to a receiver's 'observe' action. This approach could also be applied to other behavioural descriptions of the system.

*Failure mode analysis*

To succinctly identify potential failures in the system as a whole relating to interface failures between human and AI subsystems, we study the possible failure modes for each individual interaction. To facilitate such an analysis, we apply one or several systematic yet generic taxonomies of possible failure modes, which provide useful 'lenses' to view the interaction. Although each individual lens may not give a complete view of possible failure modes, together they force the analyst to confront the interaction from multiple perspectives and thereby challenge unconscious assumptions. To illustrate this, we consider an interaction in which an AI system is providing information to a human user. Two particularly useful lenses for this interaction are those of human intent and machine behaviours.

<u>Human Intent Lens</u>**:** The human intent lens contrasts the intent of the system's designer against that of a human user, and may be summarised through a **'use, misuse, abuse, disuse'** framework. For the system analyst, this presents the question: "*Is the human user intending to use the output of the AI system in the way the designer intended it to be used?*" This addresses some of the key concerns of highly-autonomous AI-enabled systems – that they are easily used beyond their intended scope of design, and this is where significant failure can occur.

<u>Machine Behaviour Lens</u>: Alternatively, we may view the same interaction from the perspective of possible failures in the machine's output. Through this lens, the analyst instead views the interaction through the following set of questions. We have expounded on each question to give a flavour of how they relate to modern AI systems, although this is not intended to be an exhaustive discussion of the many intriguing details of, and interplays between, these concepts.

| **Accuracy** | *"For an input sampled from a given distribution, what is the probability that the system produces an acceptable response?"* |
|---|---|

Accuracy can (and routinely should) be broken down across multiple facets: for example, the accuracy of the system should be considered for individual sub-classes/sub-populations and contrasted between the data used to train and validate the system and data from the expected operational environment. Here, we use the loose terms 'acceptable' and 'unacceptable' to describe the outputs produced by the system, in contrast to more standard counterparts such as 'correct' and 'incorrect'. This is to explicitly encompass systems such as AI image generators, which operate in settings where correctness in an abstract sense is an ill-defined concept, but a designer may well view certain outputs as unacceptable for their application. Related notions such as *precision* (the proportion of items assigned a specific label by an automated system which should actually have that label) and *recall* (the proportion of items with a specific label which are correctly given that label by an automated system) give further nuance to an analysis of accuracy.

| **Bias** | *"Is there persistent structure to unacceptable responses produced by the system?"* |
|---|---|

This is deliberately formulated to be agnostic as to the source of such a bias (be it the data sampling process, the structure of the training algorithm, etc.) but instead focus on how its presence may impact the overall system [Shah and Sureja, 2025]. The distinction between the concepts of accuracy and bias can be subtle, and the two are routinely analysed together in practice. Despite this, the distinction in focus between them means they are useful questions to consider separately.

| **Variability** | *"If the same input is repeatedly presented to the system, how constant is the system's response?"* |
|---|---|

Generative AI, such as language models or image generators, deliberately produce randomised outputs. Such a trait may be desirable in some scenarios and highly undesirable in others. On the other hand, future outputs from stateful systems may well be influenced by past inputs. Bayesian inference techniques have such a structure, and repeatedly presenting the same input to the system may reinforce the importance of that piece of input data to the system.

| **Stability** | *"If a small change is made to the system's input, how much does the system's output change?"* |
|---|---|

A system which reports a 'green light' when a continuous variable is above 50%, and a 'red light' otherwise is an example of a system which is unstable in this sense, because small fluctuations around 50% can dramatically change the output. Hysteresis can be introduced to remove this instability in this simple case but may not always be possible. Adversarial attacks, for example, exploit a similar structure in modern AI systems [Bruna et al., 2013] and may even be an inevitable facet [Bastounis et al., 2023, 2025]. In such systems, it is also important to consider that subtle differences in the definition of stability can have profound impacts on how the system must be designed [Sutton et al., 2024].

| **Uncertainty** | *"How does the system handle inputs with differing levels of confidence, and how does the system report the confidence level of its output?"* |
|---|---|

For example, consider a system which reads in several measurements, each with an associated measure of certainty. If the system disregards the uncertainty associated with each input, it will struggle to cope in scenarios where an uncertain input contradicts a very certain input. Likewise, a system which is unable to convey the uncertainty in its computed output can present difficulties for downstream decision-making. Techniques from the field of Uncertainty Quantification [Smith, 2013] are designed to tackle such problems, but are routinely absent from AI systems.

| **Robustness** | *"Does the system's performance degrade gracefully for inputs sampled near the edge or slightly outside the system's design domain?"* |
|---|---|

Systems incorporating Q-learning techniques from reinforcement learning provide a standard example of methods which struggle with robustness in this sense. Due to their lookup-table structure, these methods are unable to respond to inputs which were not encountered at the time of training. On a broader level, untangling the interplay between the processes of memorisation (when an AI system learns to act like a lookup table) and generalisation beyond the model's training data in deep learning systems remains a challenging open problem [Zhang et al., 2021].

As demonstrated, each of these questions invites the system analyst to examine the system in further detail, and to incorporate mitigation or prevention strategies for each. This is not intended to provide an exclusive lens, and the authors support the development of a wide variety of lenses to explore different facets of the failure of HAT interactions – such as resilience or time-dependency.

*Failure tracing*

For each failure mode, analysts should trace its upstream causes and downstream impacts using the activity diagram. Analogous to eigenvalue analysis, certain design patterns can amplify or dampen these modes as they propagate. A well-designed system aims to dampen failure modes with serious consequences. For example, an unstable system followed by a system which applies hysteresis to its input mitigates the instability problem. However, sequential unstable systems may amplify the instability throughout the remainder of the system. Though we don't offer a fixed method—since analysis is system-specific—this approach is essential to comprehensive system evaluation.

## Worked example

For illustrative purposes we have developed an example use case concerning a decision support AI used in an Air Traffic Control System. This does not describe a real Air Traffic Control system.

> "An airport has implemented a new AI support system within their Air Traffic Control System to optimise their runway scheduling. The AI makes recommendations to the Air Traffic Controllers on where to adjust the live schedule to reduce the impact of delays propagating across other flights."

The particular use case focused on has been inspired by an incident which occurred in December 2023 when an opportunity to land an aircraft earlier after it had declared a fuel emergency was missed because the Air Traffic Controller "was faced with a complex and high workload scenario" [Air Accidents Investigation Branch, 2025].

An activity diagram was created for a use case where an aircraft requests an unplanned landing at the airport. Within this the 'Determine Landing Sequence' activity was identified as a HAT task and was modelled in an OODA$^2$ Diagram, which is shown below in Figure 1. Within this activity both the human and autonomy assess the situation using available data and determine an option for where to insert the additional aircraft in the landing sequence. The human may decide to use their own plan or accept the recommendation of the autonomy. The interaction between the Human-Machine Interface (HMI) 'Recommend new landing sequence' and the human 'Observe landing sequence recommendation' actions has been highlighted.

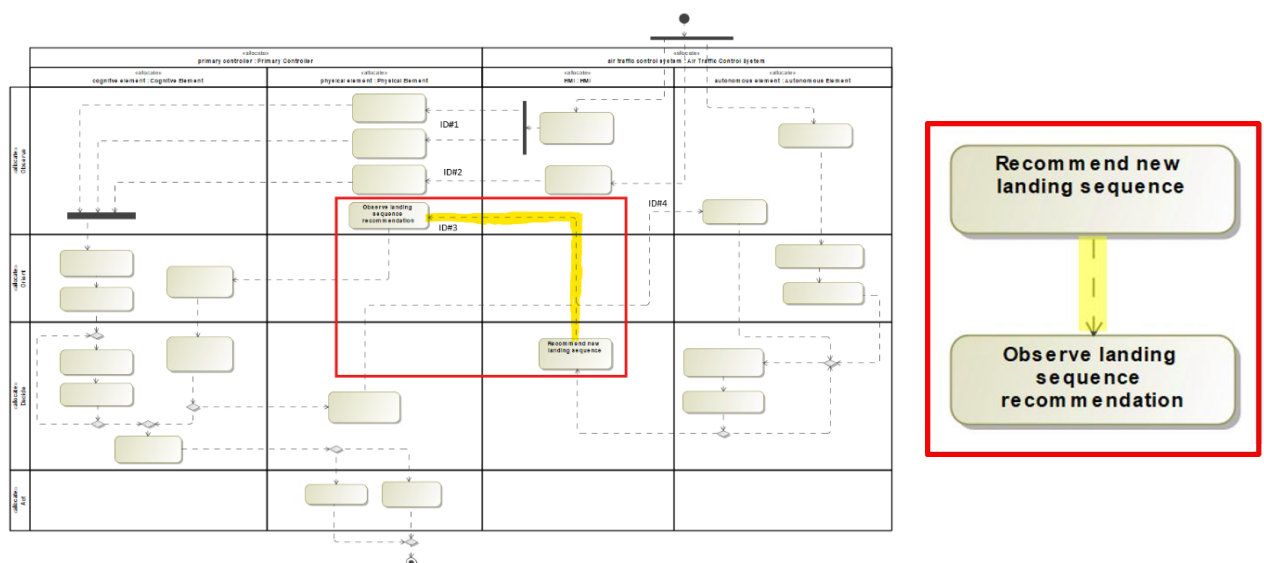

Figure 1 – example OODA$^2$ diagram for the "Determine Landing Sequence" activity

From Figure 1, we extracted each interaction between the human and machine parts and mapped the generic failure modes against them. A subset of the Failure Modes is shown below in Table 1, with Specialised Failure Modes (SFM) representing the refinement of Generic Failure Modes to be suitable for the application specific interactions.

For Interaction SFM ID#4, we have traced the failure through the system, to produce the decision pathway shown in Figure 2. Here we can see that the actions after the interaction of interest lead to the human making the decision whether to use their own judgement or use the autonomy's recommendation. Thus, the proliferation of the impact of failures through the pathway must be considered to identify potential risks. For example, the impact of not being able to understand the recommendation in a timely manner negates the benefit provided by the AI.

| I ID | SFM ID | Interaction Name | Machine Stage | Human Stage | Direction | Generic Failure Mode | Specialised Failure Mode |
|---|---|---|---|---|---|---|---|
| 1 | 1 | … | Observe | Observe | Machine->Human | … | … |
| 2 | 2 | … | Observe | Observe | Machine->Human | … | … |
| 3 | 3 | Observe Landing Sequence | Decide | Observe | Machine->Human | Autonomy output is unstable | The recommended sequence is changing frequently |
| 3 | 4 | Observe Landing Sequence | Decide | Observe | Machine->Human | Autonomy output is not understandable in a timely manner | The recommendation is incomprehensible to the operator |
| 3 | 5 | Observe Landing Sequence | Decide | Observe | Machine->Human | Autonomy output is not understandable in a timely manner | The recommendation requires too much cognition time from the operator to understand |
| 4 | 6 | … | Observe | Decide | Human->Machine | … | … |

**Table 1 – Subset of Interaction Failure Modes for "Determine Landing Sequence" activity**

Identification of risks enables second order impacts to be considered. Repeated realisation of the previously identified risk may lead the human to ignore the autonomy without trying to understand it, believing that it will offer no credible assistance. This is an example of Disuse from the human intent lens. Alternatively, the same disruption from SFM#4 could lead to the human using the results of the autonomy anyway, without being able to understand the recommendation. This is an example of Misuse from the human intent lens. When we also trace upstream through the scenario, we identify two causes that may lead to the downstream failures – data at the edge of the operational envelope leading to performance degradation (robustness error) and the generated sequence frequently altering in response to small input changes (stability error).

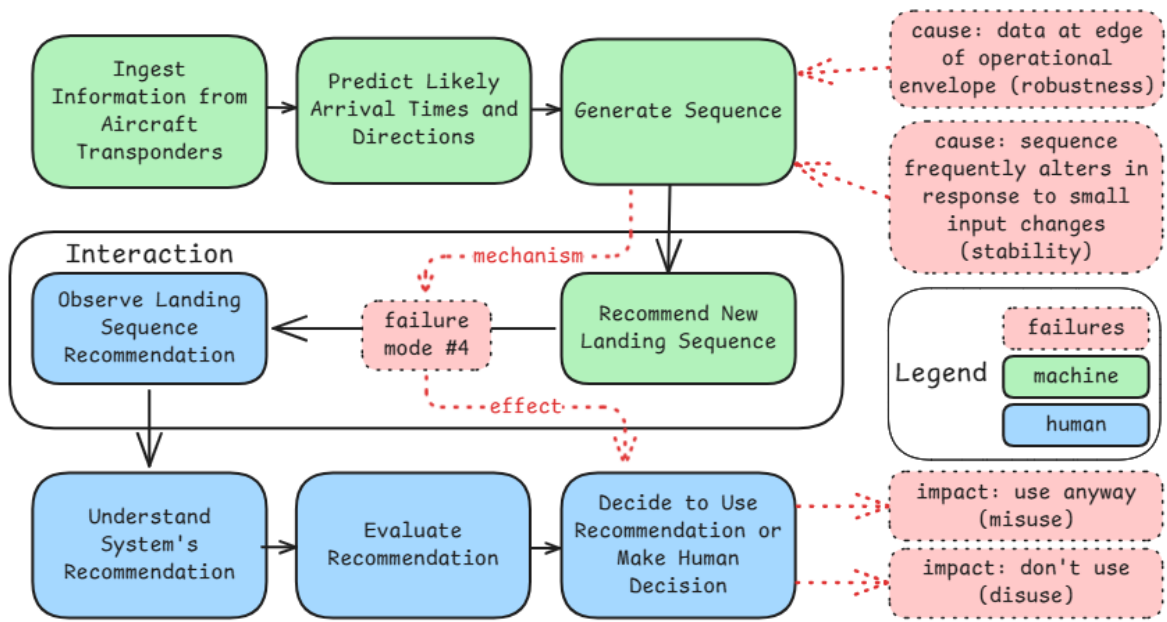

**Figure 2 – Observe Landing Sequence interaction (ID#4) pathway**

*Failure mode mitigations*

Each failure mode, alongside each failure mechanism and effect, will require different mitigations. This section gives examples of how the specific examples above could be mitigated.

Machine->Human Failure (Robustness): in the example where the failure is caused by data that exists at the edge of the operational envelope, careful consideration should be given to the definition of this envelope in the context of operational design domain (ODD) definition. Mitigations could include notification when reaching the edge of the ODD (though this could be difficult to detect), or extending the ODD to add a "safety margin" around the expected operating envelope. However, it is important to recognise that validation of ODDs is very challenging and an open research question.

Human->Machine Failure (Misuse/Disuse): there are a number of factors that could be considered in the mitigation of these failures of human intent. Important considerations could include development of trust/calibration of trust in the system (perhaps through training, but also including the importance of culture), but also consideration of human factors/comprehensibility. Additionally, mitigations could include ongoing monitoring of operators for unusual decision acceptance patterns.

## Conclusions and future development

We have presented a principled and systematic approach to studying failure modes in systems incorporating human-autonomy teams. By analysing the fundamental interactions between components which drive the overall system, our methodology provides a toolkit for undertaking more robust and earlier analysis of impact of failures. Such analysis supports early design interventions which de-risks system development and through-life operation. In addition, understanding the impact of wider system failures on the interactions between human and autonomous system components will be important towards developing resilient and robust collaborative systems, especially as the complexity of advanced AI systems continues to grow.

Further development will construct additional lenses allowing for a complete assessment across a wider range of perspectives (such as resilience, human factors, or security). We envisage that a standard library of components and behaviours could be built which are able to damp or otherwise mitigate the impacts of specific failure modes or collections of failure modes.